\newcommand{\bee}   {\begin{equation}}
\newcommand{\ene}   {\end{equation}}
\newcommand{\beqa}  {\begin{eqnarray}}
\newcommand{\enqa}  {\end{eqnarray}}
\newcommand{\beqsa} {\begin{eqnarray*}}
\newcommand{\enqsa} {\end{eqnarray*}}
\newcommand{\bea}   {\begin{array}}
\newcommand{\ena}   {\end{array}}
\newcommand{\gapproxeq}{\lower .7ex\hbox{$\;\stackrel{\textstyle >}{\sim}\;$}}
\newcommand{\lapproxeq}{\lower .7ex\hbox{$\;\stackrel{\textstyle <}{\sim}\;$}}

\def\ga{\gamma}
\def\als{{\alpha_s}}
\def\sin{\sin^2{\theta_W}}
\def\upa{\uparrow}
\def\dna{\downarrow}

\documentstyle[11pt]{article}
\begin{document}
\thispagestyle{empty}
\baselineskip=18pt

$~$
\hfill{
\begin{tabular}{l}
DSF$-$T$-$95/26\\
CPT$-$95/PE.3209\\
\end{tabular}
}

\vspace{1.5truecm}

\begin{center}

\begin{huge}
\noindent
Does one need the anomaly to \\
\vspace{.1truecm}
describe the polarized structure functions?
\end{huge}

\vspace{1.5truecm}

\noindent
{\large
F. Buccella$^{\,a}$, O. Pisanti$^{\,a,b}$, P. Santorelli$^{\,a,b}$,\\
}
\vspace{0.5truecm}

\begin{it}
$~^a$Dipartimento di Scienze Fisiche, Universit\`a ``Federico II'',\\
Pad. 19 Mostra d'Oltremare, 00195 Napoli, Italy \\
$~^b$INFN, Sezione di Napoli,\\
Pad. 20 Mostra d'Oltremare, 00195 Napoli, Italy
\end{it}

\vspace{0.5truecm}

\centerline{and}

\vspace{0.5truecm}

{\large J. Soffer}\\
\begin{it}
Centre de Physique Th\'eorique--CNRS\\
Luminy, Case 907\\
F-13288 Marseille Cedex 9, France
\end{it}
\end{center}

\vspace{2truecm}

\begin{abstract}
\noindent
The SLAC data on the $p$, $d$ and $n$ polarized structure functions
are fairly well reproduced with and without the contribution of the
anomaly. The results are compared with a previous study based mainly
on SMC data. The implications on the solution of the spin-crisis
are discussed.
\end{abstract}

\newpage

The EMC experiment \cite{emc} on polarized deep inelastic scattering
muon-proton at $<\!\! Q^2 \!\!> = 10.7$ $\,GeV^2$ led to the result
\bee
I_p = \int_0^1 g_1^p(x) dx = 0.126 \pm 0.010 \pm 0.015
{}~~~~~~\mbox{(EMC/SLAC)},
\ene
smaller than the value predicted by the Ellis and Jaffe sum rule \cite{elja}
\bee
\frac{F}{2} - \frac{D}{18} = 0.185 \pm 0.001,
\ene
and gave rise to the ``spin crisis".
An explanation of this result has been proposed in terms of the anomaly
of
the axial vector current related to the gluon polarization $\Delta G$
\cite{anom}, but, to recover the EMC result, the rather
high value of $\Delta G\sim 5$ at
$<\!\!Q^2\!\!> = 10\,GeV^2$ was necessary, as pointed out in Ref.
\cite{AltStir},
to be compensated by a large negative orbital angular momentum $<\!\!L_z\!\!>$.
This large component of the proton spin carried by the gluon and this
compensating mechanism is far from being natural, but the Bjorken
sum rule \cite{bjor} including first order QCD corrections,
\bee
I_p - I_n = \frac{1}{6} (F+D)\left(1- \frac{\als}{\pi}\right) =
\frac{1}{6} \frac{g_A}{g_V}\left(1- \frac{\als}{\pi}\right),
\ene
would be obeyed and the generally  accepted framework to describe deep
inelastic phenomena would not be spoiled.

The EMC result, assuming the validity of the Bjorken sum rule,
would imply a large negative value for
the neutron Ellis and Jaffe sum rule
\bee
I_n = \int_0^1 g_1^n(x) dx,
\ene
while the measured value at SLAC by E142 at $<\!\!Q^2\!\!> = 2\,GeV^2$
\cite{e142} with a polarized $H\!e^3$ target, is
\bee
I_n = -0.022 \pm 0.011,
\ene
in good agreement with the prediction of Ellis and Jaffe in the absence of
the anomaly contribution:
\bee
\frac{F}{3} - \frac{2}{9} D = -0.022 \pm 0.004.
\ene

Higher twist operators have been advocated \cite{EllisCarliner} to play
an important role in the interpretation of the E142 data,
which has been taken at the rather small
$<\!\!Q^2\!\!> = 2\,GeV^2$.

More recently, at CERN \cite{cern} and SLAC \cite{slac} new deep
inelastic scattering experiments have been performed with polarized
proton and deuteron targets. An analysis of the available data has been
achieved by Gehrmann and Stirling \cite{gerst} to describe the $g_1(x)$
distributions. In the framework of the explanation of the defect in the
Ellis and Jaffe sum rule for $I_p$ by means of the gluon anomaly, they
try to describe $g_1^p(x)$ and $g_1^n(x)$ in terms of the contributions
of the valence quarks, $\Delta u_v(x)$ and $\Delta d_v(x)$, and of
$\Delta G(x)$:
\beqa
g_1^p(x,Q^2) &=& \frac{2}{9} \Delta u_v(x,Q^2) + \frac{1}{18} \Delta d_v(x,Q^2)
- \frac{1}{3} \frac{\als(Q^2)}{2 \pi} \Delta G(x,Q^2), \label{e:g1p} \\
g_1^n(x,Q^2) &=& \frac{1}{18} \Delta u_v(x,Q^2) + \frac{2}{9} \Delta d_v(x,Q^2)
- \frac{1}{3} \frac{\als(Q^2)}{2 \pi} \Delta G(x,Q^2),
\enqa
with
\bee
\bea{lcl}
x \Delta u_v &=& \eta_u A_u x^{a_u} (1-x)^{b_u} (1+\ga_u
x)\vspace{.2truecm}\\
x \Delta d_v &=& \eta_d A_d x^{a_d} (1-x)^{b_d} (1+\ga_d x)
\vspace{.2truecm}\\
x \Delta G   &=& \eta_G A_G x^{a_G} (1-x)^{b_G} (1+\ga_G x)
\ena
\label{e:distr}
\ene
and $A_q = A_q(a_q,b_q,\ga_q$) (q~=~u,d,G) in such a way that
\bee
\int_0^1 \Delta q(x) dx = \eta_q.
\label{e:etaq}
\ene

At $Q_0^2 = 4\,GeV^2$, lower limit for the data considered in order to avoid
the uncertainty on the higher twist contributions, the $\eta 's$ are fixed by
\cite{gerst} ($\als(4\,GeV^2)\, = \,0.2879$):
\beqa
\eta_u(Q_0^2) &=& 2 \tilde F(Q_0^2) = 2 \left[\left(1 - \frac{3 \als(Q_0^2)}
{5 \pi}\right)F - \frac{2 \als(Q_0^2)}{15 \pi}D \right] = 0.848 \pm 0.016 \\
\nonumber \\
\eta_d(Q_0^2) &=& \tilde F(Q_0^2) - \tilde D(Q_0^2) = \left(1 -
\frac{\als(Q_0^2)}{5 \pi}\right)F - \left(1 - \frac{11 \als(Q_0^2)}{15 \pi}
\right)D \nonumber \\
& = & -0.294 \pm 0.011 \\
\nonumber \\
\eta_G(Q_0^2) &=& 1.971 \pm 0.929.
\enqa

With theoretical and phenomenological motivations they require
\bee
\bea{lcl}
a_u & = & a_d~~~~~a_G = 1 \vspace{.2truecm} \\
b_d & = & b_u + 1 = 4.64 \vspace{.2truecm}\\
\ga_u & = & \ga_d
\ena
\label{e:constr}
\ene
and they obtain a very good fit to the data, $\chi^2/N_{DF} = 0.63$,
corresponding to the parameters
\bee
\bea{lclcl}
a_u  & = & 0.46 & \pm & 0.15, \vspace{.2truecm}\\
\ga_u & = & 18.36 & \pm & 14.49, \vspace{.2truecm}\\
b_G & = & 7.44 & \pm &3.52,
\ena
\label{e:fitstir}
\ene
with the choice $\ga_G =0$.

As the authors stress, the bulk of the SLAC data falls outside their fit
region, excluding almost all the E142 neutron data. Given the large
uncertainty on the SMC deuteron data, one can conclude that the
constraint coming from $g_1^n(x)$ is not very strong and the fair
description obtained for $g_1^p(x)$ is not surprising, once $\eta_u$,
$\eta_d$, and $\eta_G$ are chosen consistently with the experimental
value found for $I_p$.

In order to test the Bjorken sum rule, which is the most appealing target of
experiments on polarized structure functions, the value of $I_n$ should
be measured and one should try to reproduce $g_1^n(x)$.

Concerning the exclusion of data below $\,Q^2\, = 4\,GeV^2$, because of
the poor theoretical knowledge on the higher twist contributions, it is
worth observing that the values of $I_p$ measured at $<\!\!Q^2\!\!> =
3\,GeV^2$ at SLAC \cite{slac}, $0.127\,\pm\,0.004\,\pm\,0.010$, and at
$<\!\!Q^2\!\!> = 10\,GeV^2$ at CERN \cite{cern},
$0.136\,\pm\,0.011\,\pm\,0.011$, are consistent, showing that for the
proton one may safely
neglect higher twist contributions.

Here we try to describe the SLAC data on proton and deuteron at
$<\!\!Q^2\!\!> = 3\,GeV^2$ and on neutron at $<\!\!Q^2\!\!> = 2\,GeV^2$
in terms of $\Delta u_v$, $\Delta d_v$, and $\Delta G$, given by
Eq.~(\ref{e:distr}), with the values of $\eta_u$ and $\eta_d$ scaled
down to $Q_0^2 = 3\,GeV^2$. We are using the values of $F$ and $D$ of
Ref. \cite{Hyp} and $\als (3\,GeV^2) \,=\,0.35\,\pm\,0.05$,
\beqa
\tilde\eta_u &=& 2 \tilde F          =  0.835 \pm 0.022 \,, \label{e:etau} \\
\tilde\eta_d &=& \tilde F - \tilde D = -0.276 \pm 0.020 \,, \label{e:etad}
\enqa
and
\bee
\tilde\eta_G =  \frac{6\pi}{\als}\left(\frac{2}{9} \eta_u + \frac{1}{18} \eta_d
 - I_p\right) = 2.3 \pm 0.7 \label{e:etaG}
\ene
to recover the measured value for $I_p= 0.127\,\pm\,0.004\,\pm\,0.010$
\cite{slac}.

As in Ref. \cite{gerst} we fix $a_u = a_d$, $\ga_u = \ga_d$, $a_G = 1$,
and further, we require
\bee
b_u\,>\,1,
{}~~~~~b_d\,>\,3,
{}~~~~~b_G\,>\,5,
{}~~~~~\ga_u = \ga_d\,>\,0,
{}~~~\mbox{and}~~~\ga_G > 0.
\ene
The lower limits for $b_u$ and $b_d$ are smaller than the values chosen
for these parameters in Ref. \cite{gerst}, as shown in our equation
(\ref{e:constr}): the smaller one for $b_u$ is taken since $u^{\upa}$
dominates at high $x$. The lower limit on $b_G$, of the order of the
corresponding exponent for G in Eq.~(\ref{e:fitstir}), has also the
purpose to prevent $\Delta G(x)$ from carrying too much proton momentum
(the momentum carried by $\Delta G$ should of course be less than the
momentum carried by $G$). In order to have a definite sign for each
spin asymmetry we do not allow values of $\ga\,<\,-1$. In the range
$(-1,0)$ there is a strong correlation between the values of $\ga$ and
$b$ (in particular $\ga\,=\,-1$ and $b$ give the same function than
$\ga\,=\,0$ and $b+1$). The large error in the
determination of $\ga_u$ in Ref. \cite{gerst} shows that this parameter
does not play a crucial role and therefore we assume all
the $\ga's$ to be positive.

To keep into account the D-state admixture for deuterium, we take
\bee
g_1^d(x) = \frac{1}{2}\left ( 1 - \frac{3}{2}\,\omega_D \right )
\left(g_1^p(x) + g_1^n(x) \right)\,,
\label{e:g1d}
\ene
where $\omega_D= 0.058$ \cite{omD}.

The best fit to the SLAC polarized structure functions of the theoretical
expressions given by Eqs.~(\ref{e:g1p})-(\ref{e:etaq}) and
(\ref{e:etau})-(\ref{e:g1d}) is found, with a $\chi^2/N_{DF} = 1.8$,
corresponding to the following values for the parameters
($a_G=1$):
\bee
\bea{cclcl}
a_u = a_d         & = & 0.57&\,\pm\, &0.03 \vspace{.2truecm}\\
b_u               & = & 2.0 &\,\pm\, &0.3 \vspace{.2truecm}\\
b_d               & = & 3.0 &\,\pm\, &0.1 \vspace{.2truecm}\\
b_G               & = & 20  &\,\pm\, &1 \vspace{.2truecm}\\
\ga_u = \ga_d     & = & 1.0 &\,\pm\, &0.8 \vspace{.2truecm}\\
\ga_G             & = & 0.0 &\,\pm\, &1.0
\ena
\label{e:fit1}
\ene
The resulting predictions are compared with SLAC data in Figs. 1, 2, 3,
showing a tendency to get large negative values at small $x$ for $g_1^d$,
$g_1^n$, just in the region where the negative contributions of $\Delta G$
are more important. The behaviour of $x\Delta u (x)$, $x\Delta d(x)$, and
$x\Delta G (x)$ is described  in Fig. 4.

If one leaves $\eta_G$ free, one gets a better fit, $\chi^2/N_{DF} = 1.0$,
with ($a_G=1$)
\bee
\bea{cclcr}
a_u = a_d  & = & 0.40&\,\pm\,&0.08 \vspace{.2truecm}\\
b_u        & = & 1.8&\,\pm\, &0.2 \vspace{.2truecm}\\
b_d        & = & 3.0&\,\pm\, &0.5 \vspace{.2truecm}\\
b_G        & = & 20 &\,\pm\, &15   \vspace{.2truecm}\\
\ga_u=\ga_d & = &  5.3 & & \displaystyle{^{+5.4}_{-2.7}} \vspace{.2truecm}\\
\ga_G      & = & 0&\,\pm\,&18 \vspace{.2truecm}\\
\eta_G/\tilde\eta_G & = & 0.48 &\pm& 0.09
\ena
\label{e:fit2}
\ene

Finally, with $\eta_G = 0$ and $\eta_u$ and $\eta_d$ free, the best fit,
with $\chi^2/N_{DF} = 1.0$,  is
obtained with
\bee
\bea{cclcl}
a_u = a_d  & = & 0.98&\,\pm\,& 0.09 \vspace{.2truecm}\\
b_u        & = & 1.8&&\displaystyle{^{+0.5}_{-0.2}} \vspace{.2truecm}\\
b_d        & = & 4.9&\,\pm\, &0.8 \vspace{.2truecm}\\
\ga_u=\ga_d& = & 0&\,\pm &3 \vspace{.2truecm}\\
\eta_u/\tilde\eta_u & = & 0.76&\,\pm\,&0.03
\vspace{.2truecm}\\
\eta_d/\tilde\eta_d & = & 0.93&\,\pm\,&0.08
\ena
\label{e:fit3}
\ene

The predictions corresponding to the values in Eqs.~(\ref{e:fit2}) and
(\ref{e:fit3}) are also compared with the SLAC data in Figs. 1, 2, 3.
The parameters given in Eqs.~(\ref{e:fit1}) and (\ref{e:fit3}) are
reported in Table \ref{t:fit} together with the parameters of reference
\cite{gerst}, hereafter referred as $Fit~A$, $Fit~B$, and $Fit~C$
respectively. It is worth stressing that $b_u$ and $b_d$ in $Fit~A$ come
out smaller than the values fixed in Ref. [10], while $b_G$ comes out
larger than the corresponding value found in Ref. [10] and the total
proton momentum carried by $\Delta G$ in $Fit~A$ is about twice smaller
than in $Fit~C$.

The values of $\eta_u$  and $\eta_d$  of $Fit~B$ support a different
interpretation for the defect in the Ellis and Jaffe sum rule for the
proton, which has been related to the defect in the Gottfried sum rule
and to the role of Pauli principle for parton distributions
\cite{bucsof}. It amounts to say that the $u^\upa$ parton, largely the
most abundant in the proton, receives less contribution from the sea
because its energy levels are almost completely occupied. Indeed, the
role of Pauli principle in parton distributions explains, in a natural
way, the dominance of $u^\upa$ at high $x$, shown by the increase
towards 1 of the asymmetry parameter $A_1^p(x)$ at high $x$ and also the
fast decrease in the same limit of the ratio $F_2^n(x)/F_2^p(x)$. This
was known since the time of the phenomenological analysis performed by
Field and Feynman \cite{feyn}, who observed that Pauli principle may be
responsible for less $u \bar u$ pairs than $d \bar d$ in the proton,
just due to the presence of two valence $u$ quarks and only one $d$
quark. The defect in the Gottfried sum rule confirms their conjecture
and it is natural to assume that the $u^\upa$ quark rather than the
$u^\dna$ quark receives less contribution from the sea. The fact that
$\eta_u$, reported in Eq.~(\ref{e:fit3}), is smaller by 25\% than the
expected value, while the reduction of $\eta_d$ is less relevant, complies
with this picture.

The importance of the role of Pauli principle in the parton
distributions therefore suggests to describe them as Fermi-Dirac
functions
\cite{fitsof,fitbuc}:
\bee
p(x) = f(x)\left[exp\left\{\frac{x-\tilde x(p)}{\bar x}\right\}+1\right]^{-1}
\ene
where $\bar x$, $\tilde x(p)$, and $f(x)$ play the role of the
``temperature", ``thermodynamical potential", depending on the flavour
and on the spin of the parton, and ``weight function" for the density of
levels in the $x$ variable.

Two models have been proposed with the common feature of relating the shape
of the distributions of the different partons to their first moments
(broader shapes for larger first moments). In one of them \cite{fitsof}
the polarized distributions are found, without additional parameters, from
the unpolarized distributions with the assumptions:
\beqa
\Delta u(x) + \Delta\bar u(x) &=& u(x) +\bar u(x) -d(x) -\bar d(x)
\label{e:defsof} \\
\Delta d(x) &=& (F-D) \left( d(x) - \bar d(x) \right) \\
\Delta \bar d(x) & = & 0
\enqa

In the second one \cite{fitbuc} the potentials of $u$ and $d$ of both
helicities
are put as parameters in order to describe unpolarized and polarized available
distributions.

Eq.~(\ref{e:defsof}) would imply, with only $u$, $d$ and their antiparticles
contributing to the polarized structure functions,
\bee
x \left[g_1^p (x) - \frac{1}{4} g_1^n(x) \right] = \frac{5}{8}
\left[F_2^p (x) - F_2^n (x) \right]\, .
\label{g1f2}
\ene

The distributions found in \cite{fitbuc}, where the nucleon unpolarized
structure functions $F_2^p(x)$, $F_2^n(x)$, $x\bar q(x)$, and $xF_3(x)$
were also described, give the following results
\bee
\bea{lcl}
\Delta u (x) & = & {\displaystyle 2.66\, x^{-0.203}\, (1-x)^{2.34}
\left(\left[exp\left\{\frac{x-1.00}{0.235}\right\} + 1\right]^{-1}  -
\left[exp\left\{\frac{x-0.123}{0.235}\right\} + 1 \right]^{-1}\right)}
\vspace{.2truecm}\\
\Delta d (x) & = & {\displaystyle 2.66\, x^{-0.203}\, (1-x)^{2.34}
\left(\left[exp\left\{\frac{x+0.068}{0.235}\right\} + 1 \right]^{-1} -
\left[exp\left\{\frac{x-0.200}{0.235}\right\} + 1 \right]^{-1}\right)}\,,
\ena
\label{e:fitbuc}
\ene
in good agreement with the Eq.~(\ref{g1f2}) and with $Fit~B$, as shown in
Fig. 5.
\footnote{
The values for $b_u$ in Eqs.~(\ref{e:fit1}), (\ref{e:fit2}) and
(\ref{e:fit3}) are smaller than the exponents fixed in Ref. \cite{gerst}
by the positivity requirement that $\Delta u (x)$ should be smaller than
parameterization assumed there for $u(x)$.\\
We think there is no problem of positivity since $\Delta u(x)$ as
obtained by Eqs.~(\ref{e:distr}) and (\ref{e:fit3}) is in good agreement,
as shown by Fig. 5, with $\Delta u(x) \, = \, u^{\upa}(x)\, -\,u^{\dna}(x)$
of Ref. \cite{fitbuc} shown in Eq.~(\ref{e:fitbuc}), with $u(x)\, =
\, u^{\upa}(x)\, +\, u^{\dna}(x)$ in good agreement with the
experimental unpolarized distributions.
}

We conclude
that the SLAC data might be fairly described in terms of $\Delta u(x)$ and
$\Delta d(x)$, with a smaller first moment for $\Delta u(x)$ than
expected. We cannot, however, conclude that the SLAC data give evidence
against the Bjorken sum rule. In fact, with $\eta_u$ and $\eta_d$
given in Eqs.~(\ref{e:etau}) and (\ref{e:etad}), it is possible to find a
good fit to the distributions by leaving $\eta_G$ free with a value $1/2$
of the rhs of Eq.~(\ref{e:etaG}).

Concerning the very important issue of the validity of the Bjorken sum
rule, it is worth stressing that the SMC data on $g_1^d(x)$ imply a
value for $I_n$ more negative than the SLAC data with polarized
$H\!e^3$, expecially from the contributions of $g_1^n(x)$ in the small
$x$ region not yet measured by SLAC. More precise measurements in this
region would be very crucial to clarify this point: indeed, two new
experiments at SLAC, E154 and E155, will be running this year and
next year with an electron beam energy of $50\,\, GeV/c$, and we hope
their results will help solving this important problem. Precise
measurements in the small $x$ region are in fact very relevant also
since the negative contribution from the gluons advocated to account for
the defect in the Ellis and Jaffe sum rule for the proton is expected to
dominate the small $x$ region.

\newpage

\begin{table}[h]
\begin{center}
TABLE I \\
\vspace{.8truecm}
\begin{small}
\begin{tabular}{|c|c|c|c|} \hline\hline
&&& \\
& $Fit~A$ & $Fit~B$  & $Fit~C$ \\
&&& \\
\hline\hline
		&               &          &    \\
$\chi^2/N_{DF}$ &   1.8         &   1.0    &  0.63         \\
		&               &                     &\\ \hline
		&                 &                   &\\
$a_u=a_d$       & $0.57 \pm 0.03$ &  $ 0.98 \pm 0.09$ &  $0.46 \pm 0.15$ \\
		&               &                     & \\
\hline
		&               &               &      \\
$b_u$           & $2.0 \pm 0.3$ & $ 1.8 ~~{\displaystyle
^{+0.5}_{-0.2}}$  &     3.64 (fixed) \\
		&               &               & \\ \hline
		&               &               & \\
$b_d$           & $3.0 \pm 0.1$ & $4.9 \pm 0.8$ & 4.64 (fixed) \\
		&               &               & \\ \hline
		&               &               & \\
$\ga_u=\ga_d$   & $1.0 \pm 0.8$ & $0 \pm 3$     &  $18.36 \pm 14.49$\\
		&               &               & \\
\hline
		&               &               & \\
$a_G$           &  1 (fixed)    &   -           &  1 (fixed) \\
		&               &               & \\
\hline
		&               &               & \\
$b_G$           & $ 20 \pm 1$   &   -           &  $ 7.44 \pm 3.52$ \\
		&               &               & \\
\hline
		&               &               & \\
$\ga_G$         &   $0 \pm 1$   &   -           &  0 (fixed) \\
		&               &               & \\
\hline
		&               &               & \\
$\eta_u/\tilde\eta_u$   &   1 (fixed)   & $0.76 \pm 0.03$ & 1 (fixed)   \\
		&               &                & \\
\hline
		&               &                & \\
$\eta_d/\tilde\eta_d$ &   1 (fixed) &  $0.93 \pm 0.08$ & 1 (fixed)      \\
		&               &               &  \\
\hline
		&               &               &  \\
$\eta_G/\tilde\eta_G$   &   1 (fixed)           &   0   (fixed)
&1 (fixed) \\
		 &              &               & \\ \hline
		 &              &               & \\
$\Delta u^{(2)}$ &   0.16       &   0.17        & 0.15 \\
		 &              &               & \\ \hline
		 &              &               & \\
$\Delta d^{(2)}$ &  $-$0.039    &   $-$0.037    & $-$  0.042\\
		 &              &               & \\ \hline
		 &              &               & \\
$\Delta G^{(2)}$ &   0.10       &   0           &  0.24 \\
	  &                              &  & \\ \hline
\end{tabular}
\end{small}
\end{center}
\caption{
We report, for a comparison, the fitted values for the parameters of the
polarized parton distributions and their first and second momenta for
the cases discussed in the paper. $Fit~C$ is the one of  Ref. [10] based
mainly on SMC data. $Fit~A$ concerns SLAC data with mainly the same
ansatz of Ref. [10], while $Fit~B$ concerns the same data assuming no
gluon anomaly contribution. }
\label{t:fit}
\end{table}

\centerline{{\large\bf Figure Captions}}

\vspace{2truecm}

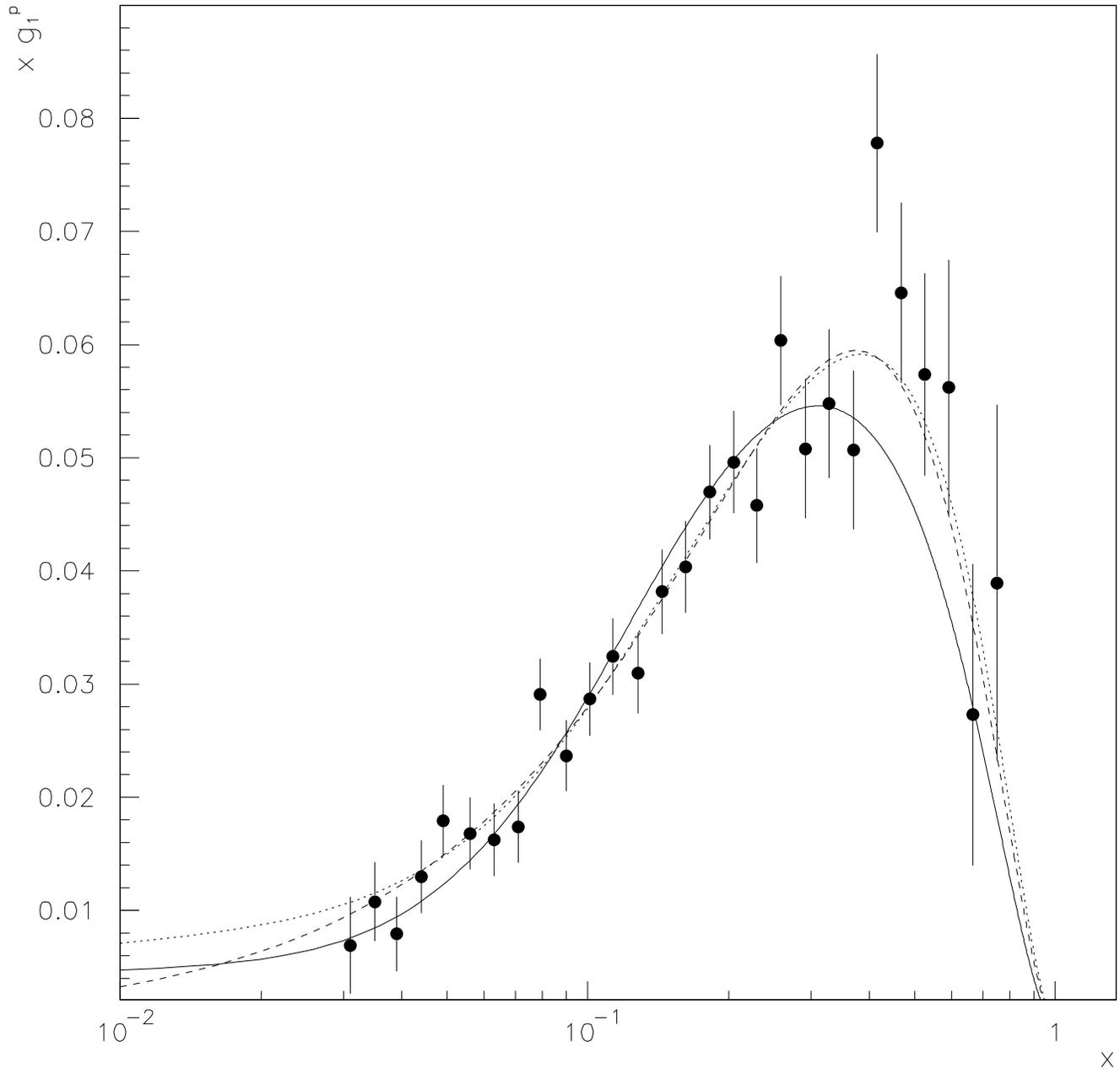
{\bf Figure 1.}
The predictions about $x\,g_1^p(x)$ corresponding to $Fit~A$
(continuous line), $Fit~B$ (dashed line) and to the values of the
parameters reported in Eq.~(\ref{e:fit2}) (dotted line) are
compared with SLAC data.

\vspace{1.5truecm}

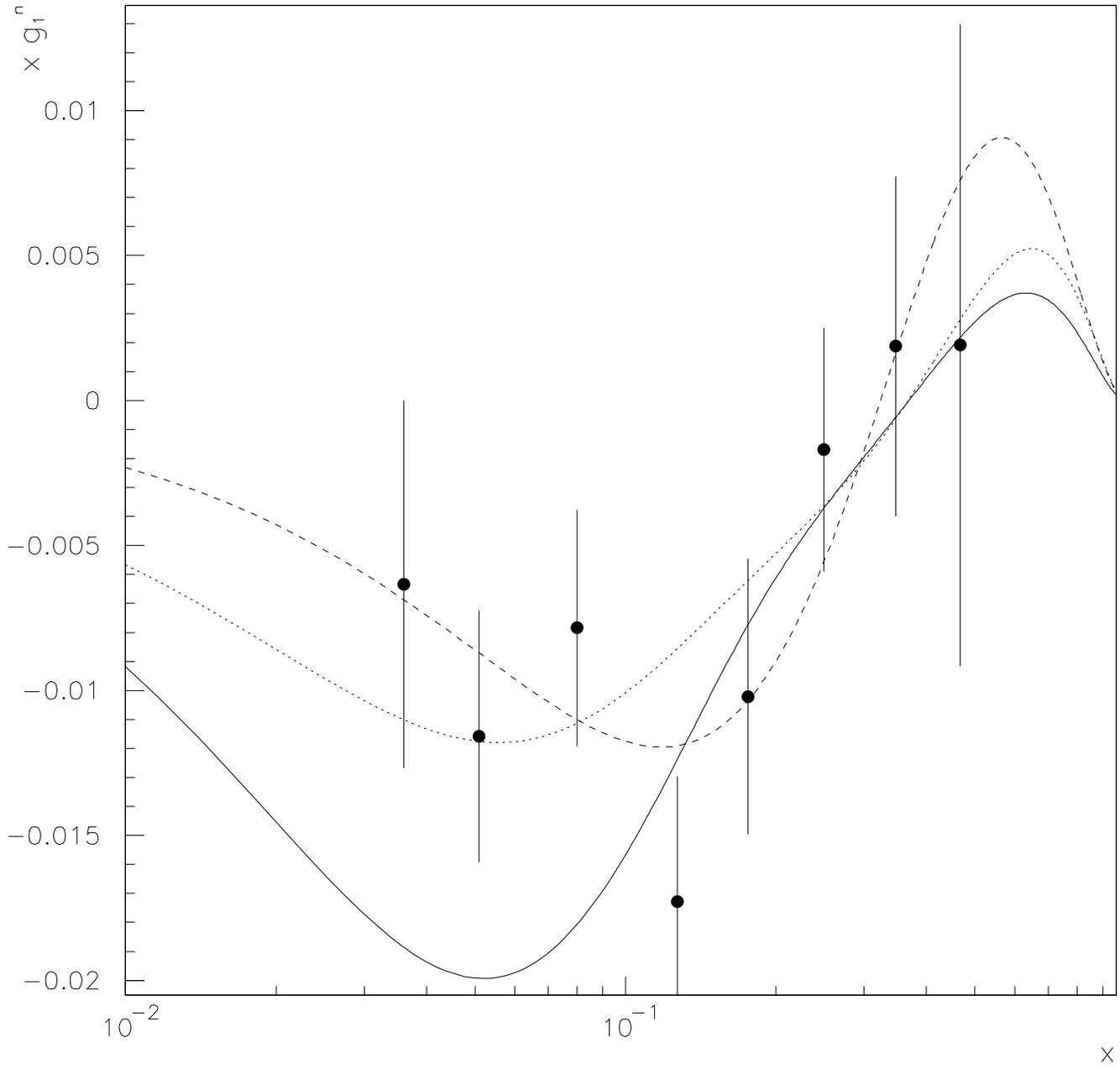
{\bf Figure 2.}
The predictions about $x\,g_1^n(x)$ corresponding to $Fit~A$ (continuous
line), $Fit~B$ (dashed line) and to the values of the parameters
reported in Eq.~(\ref{e:fit2}) (dotted line) are compared with SLAC data.

\vspace{1.5truecm}

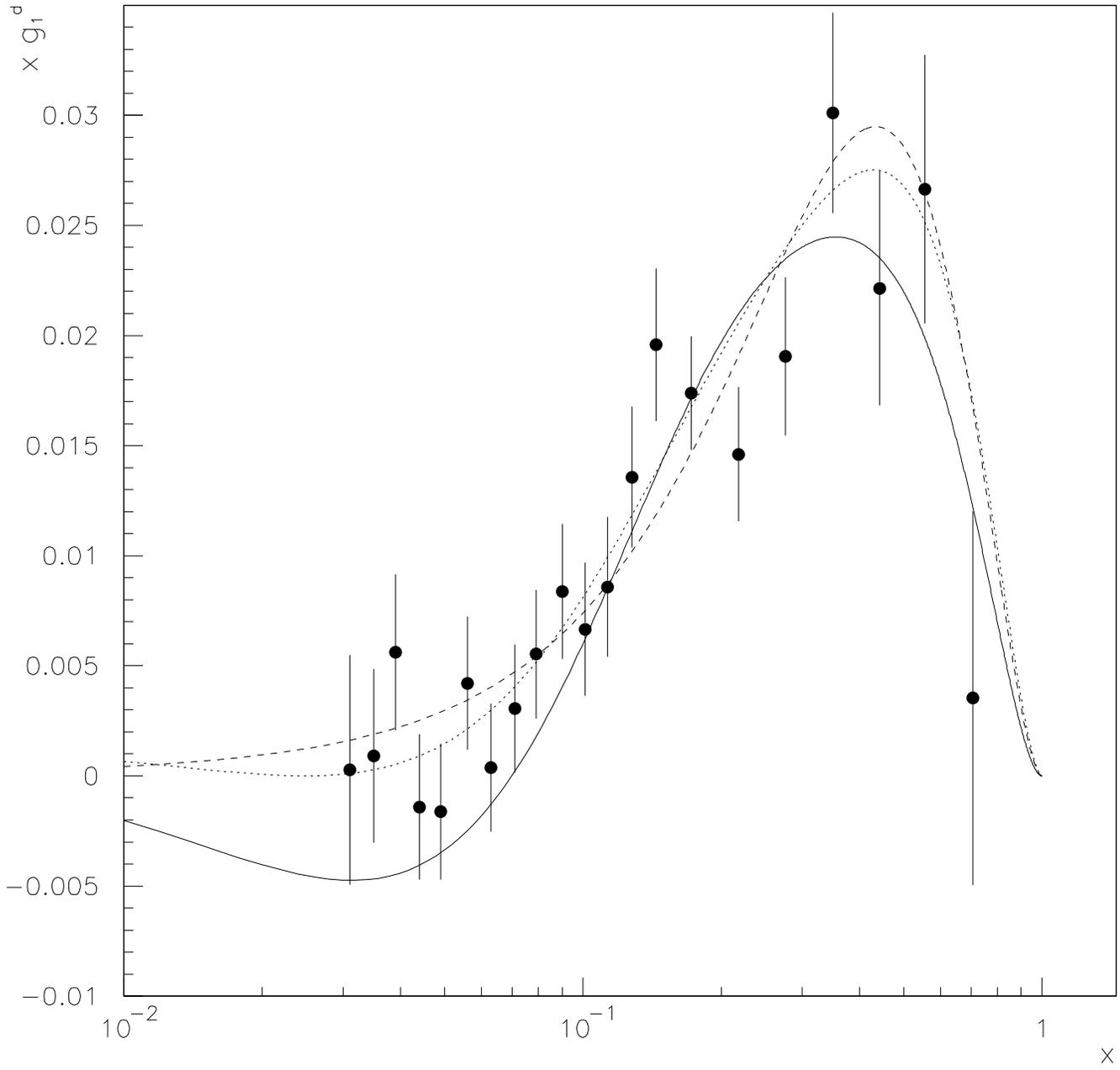
{\bf Figure 3.}
The predictions about $x\,g_1^d(x)$ corresponding to $Fit~A$ (continuous
line), $Fit~B$ (dashed line) and to the values of the parameters
reported in Eq.~(\ref{e:fit2}) (dotted line) are compared with SLAC
data.

\vspace{1.5truecm}

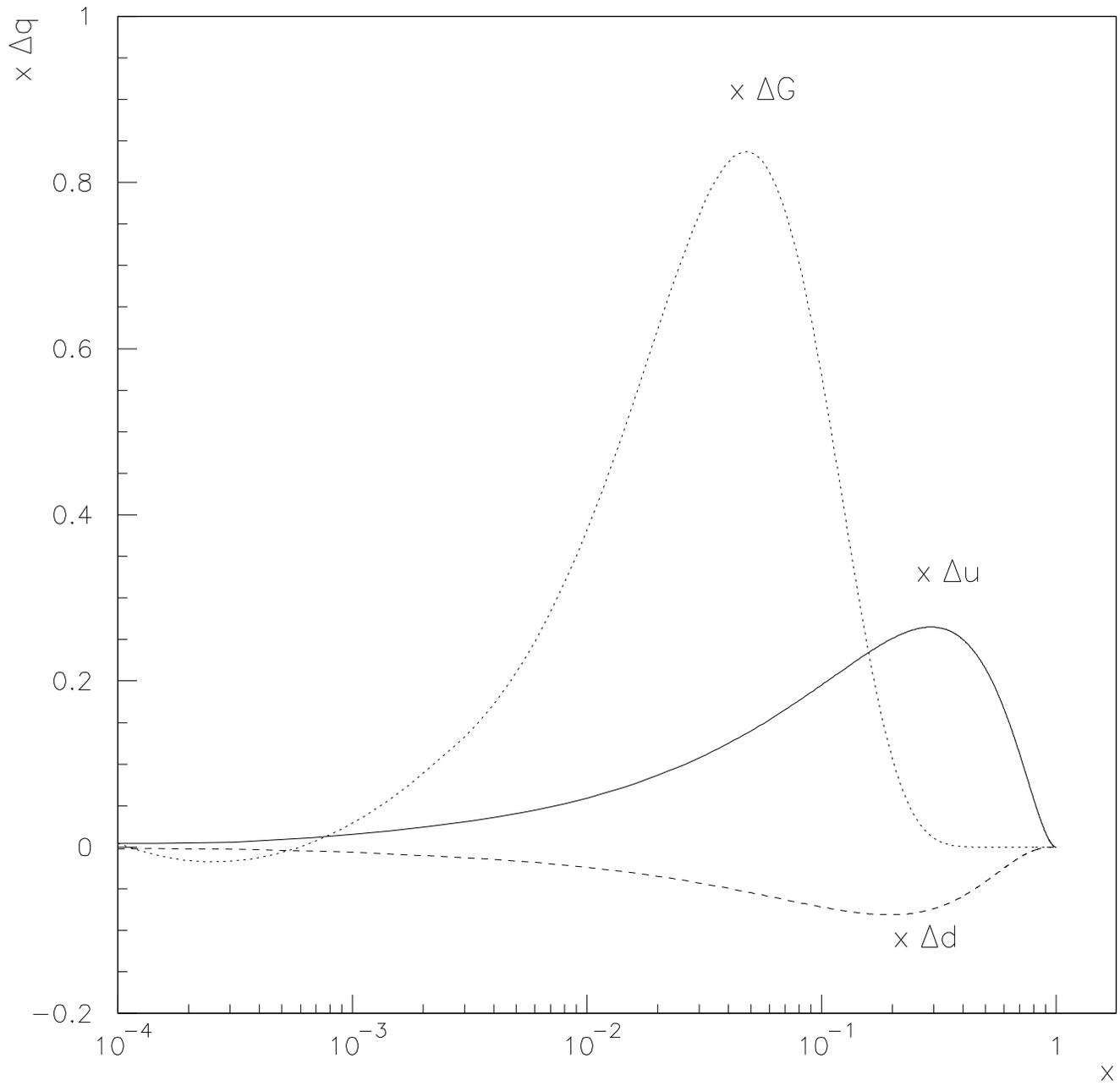
{\bf Figure 4.}
The quantities $x\,\Delta u(x)$ (continuous line), $x\,\Delta d(x)$
(dashed line) and $x\,\Delta G(x)$ (dotted line) corresponding to the
$Fit~A$ are shown.

\vspace{1.5truecm}

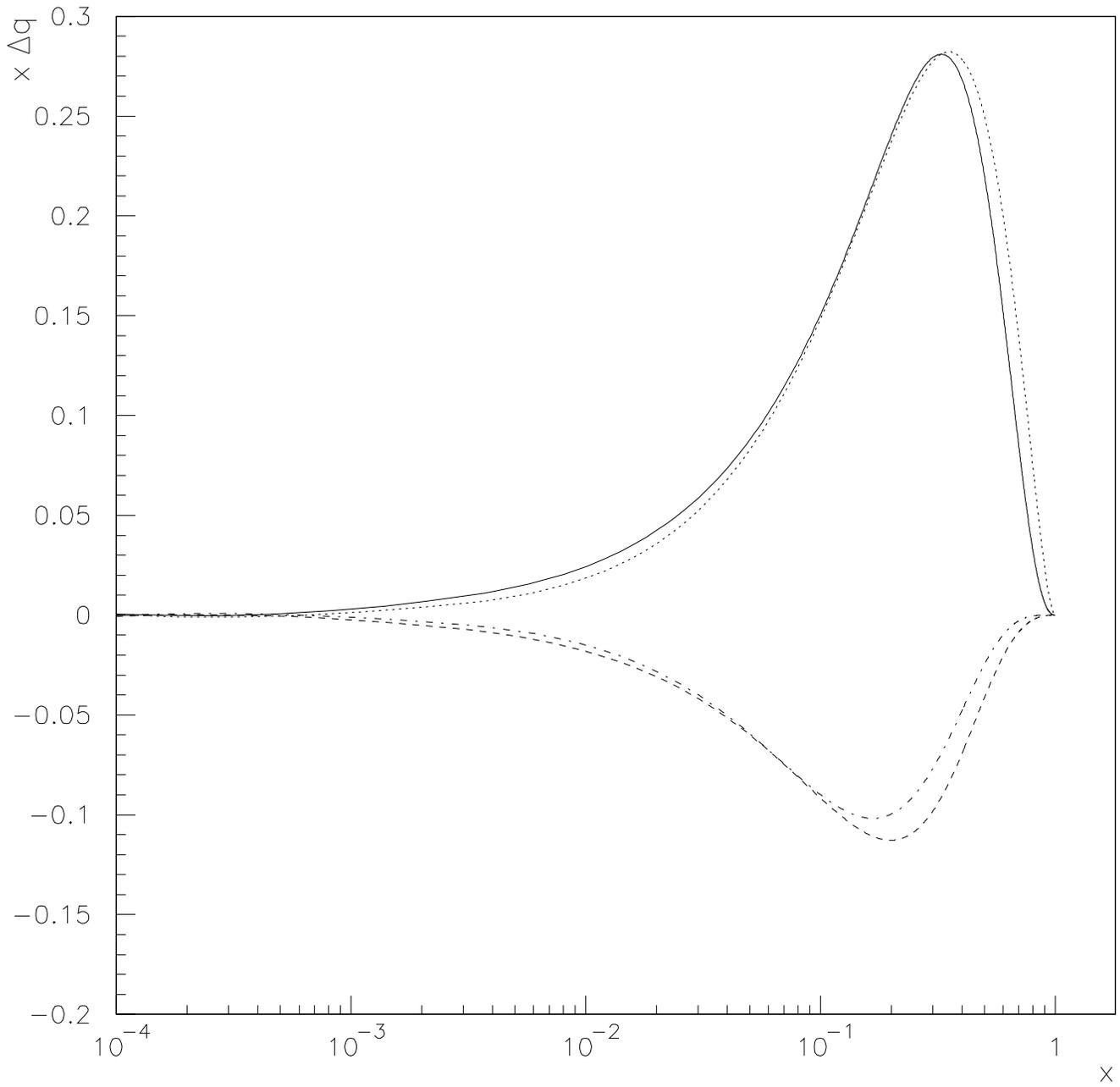
{\bf Figure 5.}
The predictions of $x\,\Delta u(x)$ (continuous line) and $x\,\Delta
d(x)$ (dashed line) obtained from Ref. [16] are compared with
$x\,\Delta u(x)$ (dotted line) and $x\,\Delta d(x)$ (dashed-dotted line)
corresponding to $Fit~B$.

\end{document}